\documentstyle[prb,aps,graphicx]{revtex}
\newcommand{\lvec}[1]{\stackrel{\small\leftarrow}{#1}}
\newcommand{\rvec}[1]{\stackrel{\small\rightarrow}{#1}}
\begin{document}

\draft
\title{Multipole response of doped $^3$He drops}
\author{Francesca Garcias, Lloren\c{c} Serra, and Montserrat Casas}
\address{Departament de F\'{\i}sica, Universitat de les Illes
Balears, E-07071 Palma de Mallorca, Spain}
\author{Manuel Barranco}
\address{Departament ECM, Facultat de F\'{\i}sica, Universitat de
Barcelona, E-08028 Barcelona, Spain}
\date{\today}
\maketitle
\begin{abstract}
The multipole response of $^3$He$_N$ drops doped with very
attractive impurities,
such as a Xe atom or an SF$_6$ molecule, has been investigated
in the framework of the Finite Range Density Functional Theory and the
Random Phase Approximation. We show that volume ($L$ = 0)
and surface ($L$ = 1, 2) modes become more fragmented, as compared with
the results obtained for pure $^3$He$_N$ drops. In
addition, the dipole mean energy goes smoothly to zero when $N$
increases, indicating that for large $N$ values
these impurities are delocalized in the bulk of the drop.
\end{abstract}


\widetext

\section{Introduction}

The study of helium drops has been the object of
extensive experimental
and theoretical investigations \cite{T90,W94,W98,T98}.
One of the goals of these studies is to understand
how various bulk
physical properties of the quantum liquid are modified in
restricted geometries.
Special mention deserves the emerging field of infrared spectroscopy
of molecules inside or attached to helium
droplets\cite{W98,T98,Gre98,Nau99} which has motivated
a great theoretical activity to determine how the molecular
moments of inertia are affected by the helium environment
\cite{Cal99,Kwo99}.

The main experimental effort has focused on the study of
pure and doped $^4$He$_N$ drops, for which a microscopic
description
of the ground state (gs) using Monte Carlo techniques
\cite{W98,Sin89,Kri90,Chi90,Bel94,Kwo96,Blu96,Bre00},
and of the elementary
excitations using an optimized variational
method\cite{Chi92,Chi95} are available.
Density functional calculations of the gs and excitation
spectrum using finite-range (FRDF) or zero-range density functionals
have been carried out, see Refs.
\onlinecite{Cas90,Dal94,Bar94,Ca95,Ga97} and references therein.
Recently, the physical appearance
of quantized vortices pinned by dopant molecules in $^4$He droplets
has been studied within the FRDF theory\cite{Dal00}.

In the case of $^3$He drops, experimental data are becoming
available \cite{Ha97}. Small $^3$He drops are difficult to
produce since a minimum number of atoms is needed to produce
a selfbound drop\cite{Pan86}, and are as difficult to observe as
$^4$He drops because they are neutral. Nevertheless, $^3$He systems
constitute the only Fermi systems capable of being observed in bulk
liquid and droplets, and for this reason they have attracted some
theoretical interest. Yet, microscopic calculations of
$^3$He droplets are scarce, and only concern the gs structure
\cite{Pan86,Gua00,Guab00}. 
A mass-formula for $^3$He drops based on an
extended Thomas-Fermi method has been proposed\cite{Cas93},
and the binding energy of
open-shell $^3$He drops has been determined by a semiempirical
shell-correction method\cite{Yan96}. The gs of small polarized
Li-$^3$He$_N$ clusters has been determined using the Path Integral
Monte Carlo Method\cite{Bor92}.
Ground state properties of $^3$He drops doped with some inert
atoms and molecular impurities have been recently studied \cite{Gar98}
within the FRDF theory, as well as
the gs structure of pure or doped mixed $^3$He-$^4$He droplets
\cite{Barb97,Pi99}. Studies of mixed droplets
are relevant in connection with the experimental results presented
in Ref. \onlinecite{Gre98}. Indeed, it is crucial to know the
composition of the first solvation shells around the impurity to
determine if the dopant molecule is in a superfluid environment
\cite{Gre98}, or to determine
whether the molecule may couple to bosonic or fermionic-type liquid
excitations which in turn determines the dissipative picture of the
molecule rotational spectrum \cite{Bab99}.

Previously quoted references indicate that there has been an enormous 
impetus in the development and application of microscopic techniques
to the description of liquid helium drops. However,
current experiments sometimes have to deal with situations that
cannot be addressed by fully microscopic methods.
We can mention, for example, the description of very large $^4$He
and $^3$He drops\cite{Har98,Har01}, or the structure of large
mixed drops already discussed. As a matter of fact, in spite of
the recent progress made in the variational description\cite{Gua00}
of small (up to 40 atoms) $^3$He droplets,
a simultaneous description of
ground state and elementary excitations of pure $^3$He droplets
has been only obtained within the density functional theory, using
either zero-range \cite{St87,Se91} or  finite-range density functionals
\cite{We92,Wei93,Ba93,Bar97}.
In these situations, density functional theory results provide a 
useful guide to the appearance of interesting physical phenomena 
obtained at the price of introducing some phenomenology.

The aim of this work is to analyze the distortions caused
by the presence of an impurity like a Xe atom or an SF$_6$ molecule
in the excitation spectrum of a $^3$He droplet. The solvation energies
of these impurities have been found to be negative\cite{Gar98}, and
this makes plausible the scenario underlying in our calculation.
We have analyzed this effect in the framework of the FRDF theory and the
Random Phase Approximation (RPA). This paper is organized as follows: In
Section II we briefly introduce the finite-range density functional
we use, and the particle-hole (ph) interaction employed in
the RPA calculations. In Section III we present results for the
volume $L=0$ and the low-multipolarity surface excitations.
A preliminary account of the dipole response has been
previously reported \cite{Ba98}.
In Section IV we draw the conclusions, and in an Appendix we present
an example of how the angular
decomposition of the ph matrix elements has been carried out.

\section{The Finite Range Density Functional and
Particle-hole Interaction}

In the framework of the density functional theory,
the ground state of $^3$He$_N$  doped drops is found by
minimizing the energy $E$ written as

\begin{equation}
E [\rho, \tau] = \int {d\vec r\; {\cal E} (\rho, \tau)}\ ,
\end{equation} \label{e1}
where\cite{Gar98}
\begin{eqnarray} \label{e2}
{\cal E} (\rho, \tau) &=&
{\hbar^2 \over 2 m^* (\vec r\,)} \left [\tau (\vec
r\,) - {\vec{j}^{\, 2} (\vec r\,) \over \rho (\vec r\,)}\right ]
\nonumber \\
&+& {1 \over 2} \
\rho(\vec r\,) \int d\vec r\,'\, V_{LJ} (\vert \vec r - \vec r\,'
\vert)\, \rho (\vec r\,')
+ {c\over 2}\ \rho^2(\vec r\,) [\bar \rho (\vec r\,)]^\gamma +
\rho (\vec r\,)\ V_{\em imp}(\vec r\,) \ .
\end{eqnarray}
The particle $\rho (\vec r\,)$, current $\vec{j}(\vec r\,)$
and kinetic energy $\tau (\vec r\,)$
densities are written in terms of the single particle (sp)
wave functions
$\phi_k (\vec r\,)$ obtained solving the Kohn-Sham (KS)
equations deduced from Eq. (\ref{e2}). For systems having an
effective mass $m^*(r)$, the inclusion of a term $\vec{j}^{\,2}/\rho$ in
Eq. (\ref{e2}) guarantees that the density functional is Galilean
invariant\cite{Eng75}. This term has no influence on the ground state
of time-reversal invariant, spin-saturated droplets, and for this
reason it is usually omitted. However, its contribution to the
ph interaction for systems like the present
one, in which the impurity is treated as an external field
breaking the translational invariance of the system, cannot be
neglected.

In the above expression $V_{LJ}(\vert \vec r -
\vec r\,'\vert)$ is the Lennard-Jones interatomic potential screened at
short distances
\begin{equation}\label{vlj}
V_{LJ}(r) =
\left\{
\begin{array}{lr}
4 \epsilon \left[ \left( \sigma/r \right)^{12} -
\left(\sigma/r\right)^6 \right] &
\quad\rm{if} \ \ r \geq \sigma \\
b_{LJ} \left[1 - \left( r/\sigma \right)^8 \right]
& \rm{otherwise} \ , \end{array}
\right.
\end{equation}
and the averaged density $\bar \rho (\vec r\,)$ is defined as
\begin{equation} \label{rhobar}
\bar \rho (\vec r\,) =\int d \vec r\,' \rho (\vec r\,')  W (\vert
\vec r - \vec r\,'\vert)
\end{equation}
with
\begin{equation} \label{w}
W(r) = \left\{
\begin{array}{cr}
3/(4 \pi \sigma^3) &\quad\rm{if} \ r < \sigma \\
0 & \rm{otherwise} \ . \end{array}
\right.
\end{equation}
The effective mass $m^*$ is parametrized as
$m^* = m (1-\bar \rho/ \rho_c)^{-2}$.
The set of coefficients entering the definition of
${\cal E}(\rho,\tau)$ can be found in Table I of Ref.
\onlinecite{Gar98}.
$V_{imp}$ is the helium-impurity potential taken
from Ref. \onlinecite{Tan86} in the case of Xe, and from
Ref. \onlinecite{Pa84} in
the case of SF$_6$, in its spherically averaged version.
In both cases we have assumed that the impurity is an object of
infinite mass located at the coordinate origin.

The distortion of the ground state structure of $^3$He drops due to
the presence of impurities has been described in detail in Ref.
\onlinecite{Gar98}.
To analyze the multipole excitations induced by an external
field that couples to the particle density of the drop,
we have used the time-dependent version of the density functional
theory.
For sufficiently weak external fields the response can be treated
linearly within the RPA. In this approximation the elementary
excitations of the system are described in terms of correlated
ph transitions. The amplitude of a particular
excited state in the basis of a discrete space of ph transitions is
obtained by diagonalizing the Hamiltonian $H = H_0 + V_{ph}$, which is
the sum of the KS Hamiltonian $H_0$ plus the  ph interaction
$V_{ph}$. This is done solving the RPA equation \cite{Bla86,Neg88}

\begin{equation} \label{rpa}
\left(\begin{array}{cc}
A & B \\
- B^* & - A^* \\
\end{array}\right)
\left(\begin{array}{cc}
X^{(\lambda)} \\
Y^{(\lambda)} \\
\end{array}\right) = \omega_\lambda
\left(\begin{array}{cc}
X^{(\lambda)} \\
Y^{(\lambda)}\\
\end{array}\right) \ ,
\end{equation}
where the matrices $A$ and $B$ are written in terms of
matrix elements of the interaction between the ph
pairs that can be coupled to have the desired angular momentum.

Writing the particle, current and kinetic energy densities
in terms of the sp basis $\phi_k(\vec r\,)$
and the occupation numbers $p_{kl}$
\begin{equation} \label{rho}
\rho(\vec r\,)= \sum_{kl} \phi_k^*(\vec r\,)\, p_{kl}\,
\phi_l (\vec r\,)
\end{equation}
\begin{equation}\label{current1}
\vec{j} (\vec r\,) = {1 \over 2i} \sum_{kl} (\rvec \nabla - \rvec
\nabla')\, \phi_k^*(\vec r\,')\, p_{kl}\, \phi_l (\vec r\,)\,
\vert_{\vec r = \vec r\,'}
\end{equation}
\begin{equation} \label{tau}
\tau(\vec r\,)= \sum_{kl} \vec \nabla \vec \nabla'\,
\phi_k^*(\vec r\,)\,
p_{kl}\, \phi_l (\vec r\,') \vert_{\vec r = \vec r\,'} \ ,
\end{equation}
the ph interaction is obtained\cite{Bla86,Mig67}
from the second variation of the energy functional
with respect to the occupation numbers:

\begin{equation} \label{vph1}
V_{ijkl} \equiv \langle ij \vert V_{ph} (\vec r_1, \vec r_2)
\vert kl\rangle = {\delta^2 E
 \over \delta p_{ik} \delta p_{jl}} \ .
\end{equation}
If $m^* = m$ in the density functional (a situation which we
indicate with the notation $E = E [\rho]$), the second variation of
the energy with respect to the occupation numbers taken at the ground
state straightforwardly provides the ph interaction
$V_{ph}$. This variation can be obtained as

\begin{eqnarray}\label{vph2}
{\delta^2 E [\rho] \over \delta p_{ik} \delta p_{jl}} &=&
\int d \vec r_1 d \vec
r_2 \left({\delta^2 E [\rho] \over {\delta \rho {(\vec r_2)}}
\delta
\rho {(\vec r_1)}} \right)_{gs} {\delta \rho (\vec r_1) \over
\delta p_{ik}} {\delta \rho (\vec r_2) \over \delta p_{jl}}
\\
\nonumber
&=& \int d \vec r_1 d \vec r_2\,
\phi_i^* (\vec r_1)\, \phi_j^*(\vec r_2)\,
\left({\delta^2 E [\rho] \over \delta \rho
(\vec r_2) \delta \rho (\vec r_1)} \right)_{gs}\,
\phi_k(\vec r_1)\,
\phi_l(\vec r_2)\ ,
\end{eqnarray}
and comparing with Eq. (\ref{vph1}) it results
\begin{equation}
V_{ph} (\vec r_1, \vec r_2)=\left ({\delta^2 E [\rho] \over \delta
\rho (\vec r_2) \delta \rho (\vec r_1)}\right)_{gs}  \ .
\label{vph3}
\end{equation}

The presence of a position-dependent effective mass in the
functional introduces a velocity dependence
in the ph interaction. In this case Eq. (\ref{vph1}) becomes
\begin{eqnarray} \label{vph4}
V_{ijkl} = \int d \vec r_1 d \vec r_2
& \left( \rule{0mm}{6mm}\right.
{\displaystyle\delta^2 E \over\displaystyle\delta \rho (\vec r_1)
\delta \rho (\vec r_2)} &
{\delta \rho (\vec r_1) \over \delta p_{ik}}
{\delta \rho (\vec r_2) \over \delta p_{jl}} \nonumber \\
&+{\displaystyle\delta^2 E \over\displaystyle
\delta \rho (\vec r_1) \delta \tau (\vec
r_2)}&  {\delta \rho(\vec r_1) \over \delta p_{ik}}
{\delta \tau (\vec r_2) \over \delta p_{jl}}  \nonumber \\
&+{\displaystyle\delta^2 E \over\displaystyle
\delta \tau (\vec r_1) \delta \rho (\vec
r_2)}& {\delta \tau (\vec r_1) \over \delta p_{ik}} {\delta \rho
(\vec r_2) \over \delta p_{jl}}  \nonumber \\
&+ {\displaystyle\delta^2 E \over\displaystyle
\delta \tau (\vec r_1) \delta \tau (\vec
r_2)}&  {\delta \tau (\vec r_1) \over \delta p_{ik}} {\delta \tau
(\vec r_2) \over \delta p_{jl}}  \nonumber \\
&+ {\displaystyle\delta^2 E \over\displaystyle
\delta j^{(\alpha)} (\vec r_1)  \delta j^{(\alpha)} (\vec
r_2)}& {\delta j^{(\alpha)} (\vec r_1) \over \delta p_{ik}}
{\delta j^{(\alpha)} (\vec r_2) \over \delta p_{jl}}
\left. \rule{0mm}{6mm} \right)
\end{eqnarray}
with

\begin{eqnarray} \label{current2}
{\delta \tau (\vec r\,) \over \delta p_{kl}} &=&\rvec {\nabla}
\phi_k^* (\vec r\,) \rvec {\nabla} \phi_l (\vec r\,) \ , \nonumber \\
{\delta  j^{(\alpha)}(\vec r_1) \over \delta p_{kl}} &=& {1 \over 2i}
\left[ \phi_k^*(\vec {r_1})\, ( \rvec {\nabla}_1 - \lvec {\nabla}_1)\,
\phi_l (\vec r_2) \right] _{\vec r_1 = \vec r_2} \ .
\end{eqnarray}
The arrow on the gradient operators indicates whether they act on
the left or on the right.
The terms arising from current derivatives are essential to fulfill
the Thomas-Reiche-Kuhn (or energy-weighted) sum rule.
The contribution of current terms to the ph interaction is

\begin{eqnarray} \label{current3}
V_{ijkl} &=& \int d \vec r_{1} d \vec r_{2}\, \left( {\delta^{2} E \over
\delta j^{(\alpha)} (\vec r_{1}) \delta j^{(\alpha)} (\vec
r_{2})} \right) {\delta j^{(\alpha)} (\vec r_{1}) \over \delta p_{ik}}
{\delta j^{(\alpha)} (\vec r_{2}) \over \delta p_{jl}} \\
&=& {- \hbar^{2}\over 4m} \int d \vec {r_{1}} d \vec {r_{2}}\,
\phi^{*}_{i} (\vec r_{1})\, \phi^{*}_{j}(\vec {r_2})
(\rvec \nabla_{1} - \lvec\nabla_{1})
\left({\delta^{2} E \over \delta j^{(\alpha)} (\vec r_{1})
\delta j^{(\alpha)}(\vec r_{2})} \right)
(\rvec {\nabla}_{2} - \lvec {\nabla}_{2})\,
\phi_{k}(\vec r_{1})\,\phi_{l}(\vec r_{2})\ , \nonumber
\end{eqnarray}
where a sum over the three components $\alpha$ is assumed, and the
gradients only act on the sp wave functions. This expression
coincides with the back-flow contribution to the
ph interaction for $^{4}$He drops
\cite{Ca95}.

Particularizing to the density functional Eq. (\ref{e2}), accounting for
Eqs. (\ref{current2}), (\ref{current3}) and  that
${\displaystyle\delta^{2} E \over\displaystyle
\delta \tau (\vec r_{1}) \delta \tau (\vec
r_{2})} =0$, Eq. (\ref{vph4}) gives for the ph interaction
\begin{eqnarray} \label{vpht}
V_{ph}(\vec r_{1}, \vec r_{2}) &=& V_{LJ}(\vert \vec r_{1} - \vec r_{2}
\vert) + c \, [\bar \rho(\vec r_{1})]^{\gamma}\,
\delta(\vec r_{1} - \vec r_{2}) \nonumber \\
&+& c \, \gamma \left\{ \rho (\vec r_{1}) [\bar {\rho}
(\vec r_{1})]^{(\gamma-1)} + \rho (\vec r_{2})
[\bar \rho (\vec r_{2})]^{(\gamma - 1)} \right\}\, W (|\vec r_{1} - \vec r_{2}|)
\nonumber \\
&+& \frac{1}{2} \, \gamma \, (\gamma - 1) \, c
\int d \vec r\, \rho^{2} (\vec r\,)\, [\bar
\rho (\vec r\,)]^{(\gamma - 2)}\, W (|\vec r_{1} - \vec r\,|)\,
W (|\vec r_{2} - \vec r\,|)
\nonumber \\
&+& \int d \vec r\, \,{\hbar^{2} \over 2m}\,\, {2 \over \rho_c^{2}}\,\,
\tau(\vec r\,)\,\, W (|\vec {r_{1}} - \vec r\,|)\, W (|\vec r_{2} - \vec r\,|)\,
\nonumber \\
&+& \lvec {\nabla}_{1}\, f (\vec {r_{1}})\,
W (|\vec {r_{1}} - \vec {r_{2}}|)\, \rvec
\nabla_{1} + \lvec \nabla_{2}\, f (\vec {r_{2}})\, W (|\vec {r_{1}} - \vec
{r_{2}}|) \rvec {\nabla}_{2} \nonumber \\
&+& g (\vec r_{1})\, \delta(\vec r_{1} - \vec r_{2})\, (\rvec \nabla_{1} -
\lvec \nabla_{1})\, (\rvec \nabla_{2} - \lvec \nabla_{2})
\end{eqnarray}
with
\begin{equation} \label{f1}
f (\vec {r_i}\,) = {\hbar^{2} \over 2m} \left({2 \over \rho^{2}_{c}} \ \bar
\rho (\vec r_i\,) - {2 \over \rho_{c}} \right)
\end{equation}
and
\begin{equation} \label{g1}
g (\vec r\,) = {\hbar^{2} \over 4m} \left [\left
({\bar \rho (\vec r\,)
\over \rho_{c}} \right) ^{2} - 2 \ {\bar \rho (\vec r\,) \over \rho_{c}}
\right] {1 \over \rho (\vec {r}\,)} \ .
\end{equation}

Equation (\ref{vpht}) shows that in addition to the Lennard-Jones
potential $V_{LJ}$, Eq. (\ref{vlj}), the ph interaction  has
finite-range terms, velocity dependent components, and other terms which
combine both finite-range and velocity dependence through the presence of
gradient operators.

The next task is to calculate the matrix elements in the ph basis.
This is greatly simplified in the case of droplets with a magic number
of $^3$He atoms, the only droplets studied here. In this case, the mean
field is spherically symmetric and the angular part of the sp
wave functions is a spherical harmonic. Performing a
multipole expansion of the ph  interaction, the sum
over third components can be done and only radial integrals
remain to be numerically computed (see the Appendix for details).
This allows one to compute the RPA matrices $A$ and $B$. After
diagonalizing Eq. (\ref{rpa}), the strength function from the
gs $|0\rangle$ to the set of excited states $\{|n\rangle\}$ (with excitation
energies $\{\omega_{n}\}$) is obtained as
\begin{equation} \label{se}
S(\omega) = \sum_{n} \delta (\omega-\omega_{n})\, \vert \langle n
\vert Q_L \vert 0 \rangle \vert^2 \ ,
\end{equation}
where $Q_{L}$ is the excitation operator for which we have made the
natural choices $Q_{0} = \sum_{i} r_{i}^{2}$ for $L=0$ (volume mode),
and $Q_{L} = \sum_{i} r^{L}_{i} Y_{L0} (\Omega_{i})$ for $L=1, 2$
(surface modes).

The transition matrix element of the strength function is obtained in
terms of the solutions of RPA equation and the sp
radial wave functions $u_{i}$ defined in the Appendix.
The explicit expression for surface modes is given by
\begin{equation}
\langle 0 \vert Q_{L}\vert n\rangle = {1 \over \sqrt{2L+1}} \sum_{mi}
(X^{(n)}_{mi} - Y^{(n)}_{mi})\, \langle u_{m} \vert  r^{L}
\vert u_{i} \rangle
\langle\ell_m\vert\vert Y_L \vert\vert \ell_i\rangle \,
 \ ,
\end{equation}
where
$\langle\ell_m\vert\vert Y_L \vert\vert \ell_i\rangle$
is the
angular reduced matrix element of the excitation operator\cite{Con80}.
The corresponding expression for the monopole mode is
\begin{equation}
\langle 0 \vert Q_{0}\vert n\rangle = \sum_{mi}
(X^{(n)}_{mi} - Y^{(n)}_{mi})\, \langle u_{m} \vert  r^{2}
\vert u_{i} \rangle\,
\delta_{\ell_m\ell_i}\,
\ .
\end{equation}

The transition (also called induced) densities for the operator
$Q_L= \sum_{i} r^{L}_{i} Y_{L0} (\Omega_{i})$ that
causes surface excitations are obtained as
\begin{equation}
\rho_{n0}(r) = {1 \over \sqrt{2L+1}} \sum_{mi}
(X^{(n)}_{mi} - Y^{(n)}_{mi})\,
\langle\ell_m\vert\vert Y_L \vert\vert \ell_i\rangle\,
{u_m(r)u_i(r)\over r^2} \,\,\, ,
\end{equation}
and
$\langle 0 \vert Q_{L}\vert n\rangle = \int dr\, r^{2+L} \rho_{n0}(r)$.
The corresponding induced densities for the monopole mode are given by
\begin{equation}
\rho_{n0}(r) = \sum_{mi}
(X^{(n)}_{mi} - Y^{(n)}_{mi})\,
{u_m(r)u_i(r)\over r^2} \,
\delta_{\ell_m\ell_i}\, ,
\end{equation}
and
$\langle 0 \vert Q_{0}\vert n\rangle = \int dr\, r^{4} \rho_{n0}(r)$.

Obviously, the dimension of matrices $A$ and $B$ depends on how many
particle-hole pairs $mi$ are taken after discretizing the continuum.
We have included enough sp states so that the
Thomas-Reiche-Kuhn sum rule is satisfied within 98$\%$. We have also
checked that for pure drops the dipole mode is at zero energy
due to the translational invariance of the system.

To finish this Section, we would like to recall
that originally, density
functionals for liquid $^3$He were obtained from a
contact, velocity-dependent $^3$He-$^3$He effective 
interaction \cite{Str85} that made it rather simple to
evaluate the contribution of direct {\em and} exchange
terms to the total energy and to the ph interaction.
Later on, a finite-range component was added to the
contact interaction to improve its properties 
at finite momentum. This is the origin of the 
screened Lennard-Jones
potential\cite{Dup90,Gar92}, which takes care of two major
characteristics of the interatomic potential the
original effective He-He interaction lacked, namely the
hard core repulsion at short distances, and the
asymptotic $r^{-6}$ behavior. Thus, exchange effects,
which are known to be large in liquid $^3$He, are
phenomenologically accounted for in the density functional
through the effective parameters entering its definition.

\section{Results}

\subsection{Monopole mode}

\begin{figure}[h]
\includegraphics{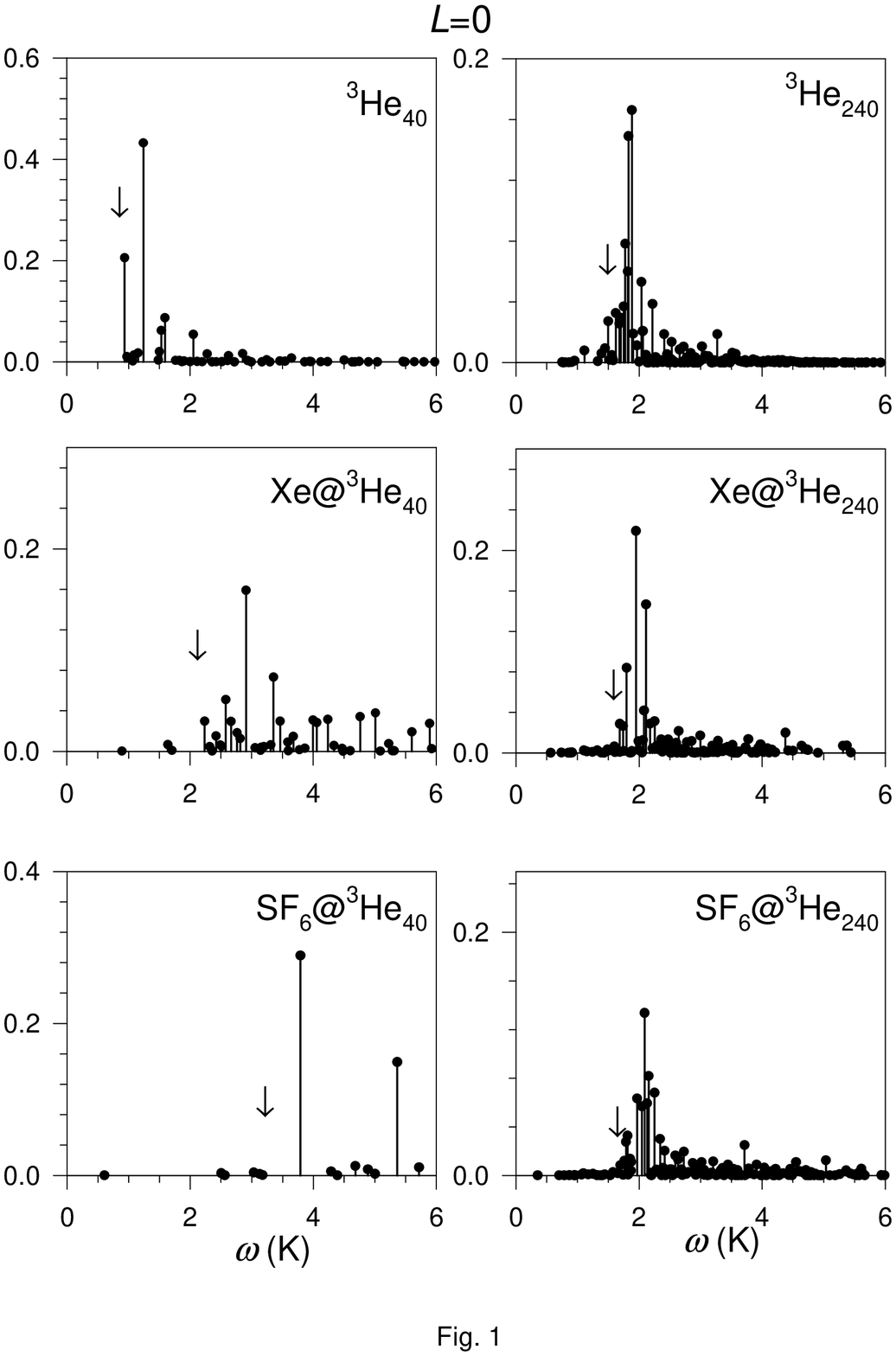}
\vspace*{19cm}
\caption{
Comparison between the monopole spectrum of pure and doped drops.
Each excited state is represented by a vertical stick whose height gives
its fractional contribution to the energy weighted sum rule.
The arrows indicate the position of the atom
emission threshold in each case.}
\label{fig1}
\end{figure}

\begin{figure}[h]
\includegraphics{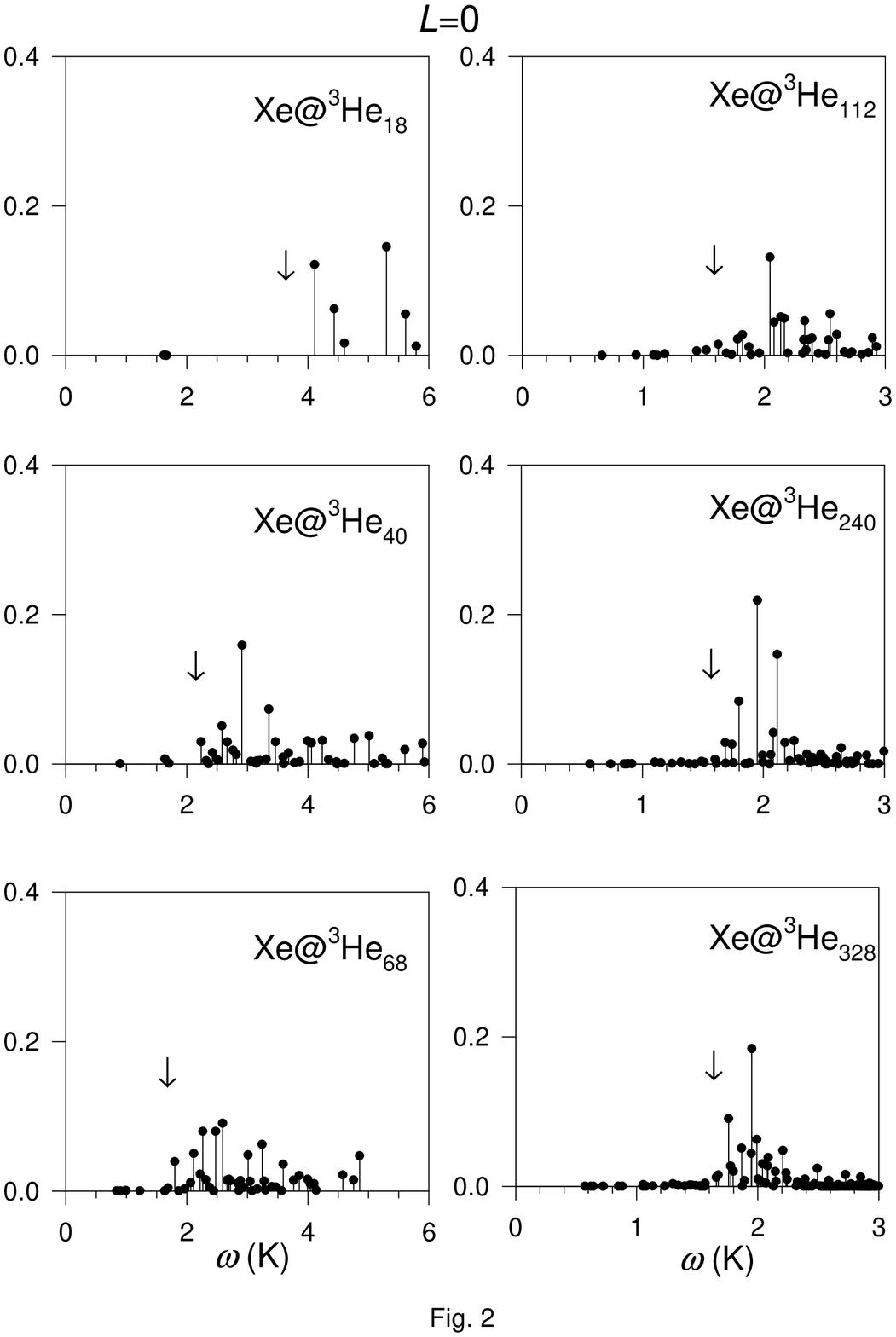}
\vspace*{18cm}
\caption{Same as Fig. \ref{fig1} for the monopole spectrum of drops doped
with Xe.}
\label{fig2}
\end{figure}

\begin{figure}[h]
\includegraphics{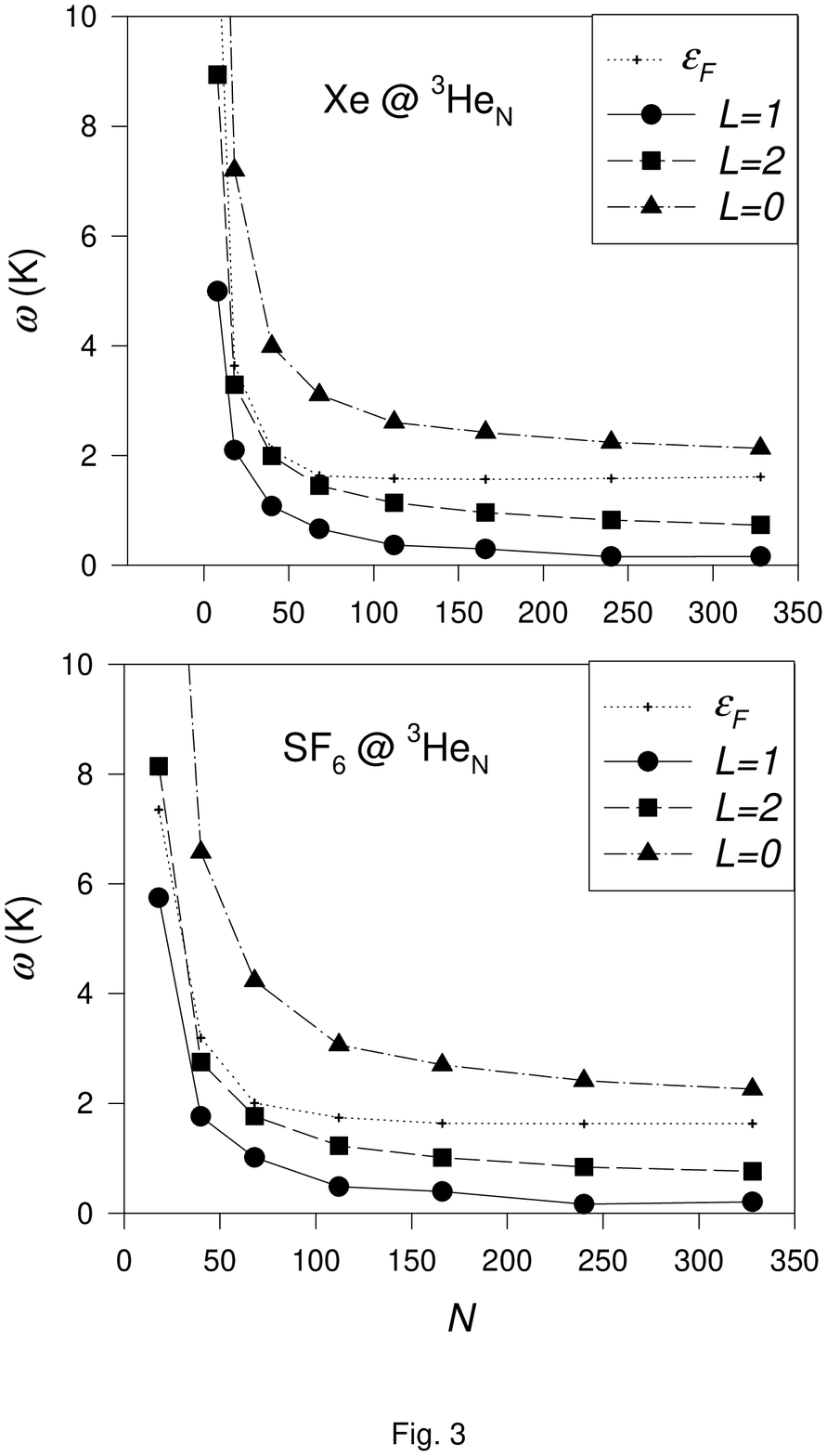}
\vspace*{18cm}
\caption{
Mean excitation energies $\bar \omega$ (K)
and chemical potential changed of sign $\varepsilon_F$
as a function of $N$ for $^3$He$_N$ drops doped with Xe and SF$_{6}$.
The lines have been drawn to guide the eye.}
\label{fig5}
\end{figure}

Figure \ref{fig1} shows a comparison between the
monopole (`breathing mode') spectrum of pure and doped drops.
It is  seen that the presence of the
impurity increases the fragmentation of the spectra in the high energy
region. This effect is more important for small clusters and more
attractive impurities.
In both cases of pure and doped drops, the mean energy defined as
$\bar{\omega} = \sum_j \omega_j S(\omega_j)/ \sum_j S(\omega_j)$ lies
above the atom emission threshold (Fermi energy changed of sign)
and decreases as the
number of $^{3}$He atoms of the drop increases (see Figs. 
\ref{fig2} and \ref{fig5}).
It is worthwhile to recall that for
pure $^4$He$_N$ droplets, except for rather small
$N$ values the monopole strength is in the discrete
region of the spectrum \cite{Chi92,Cas90,Bar94}, and that
the presence of a Xe or SF$_6$ impurity also increases
the fragmentation of the spectrum; for small drops the monopole
strength lies in the continuum region\cite{Bar94,Ga97}.

\begin{figure}[h]
\includegraphics{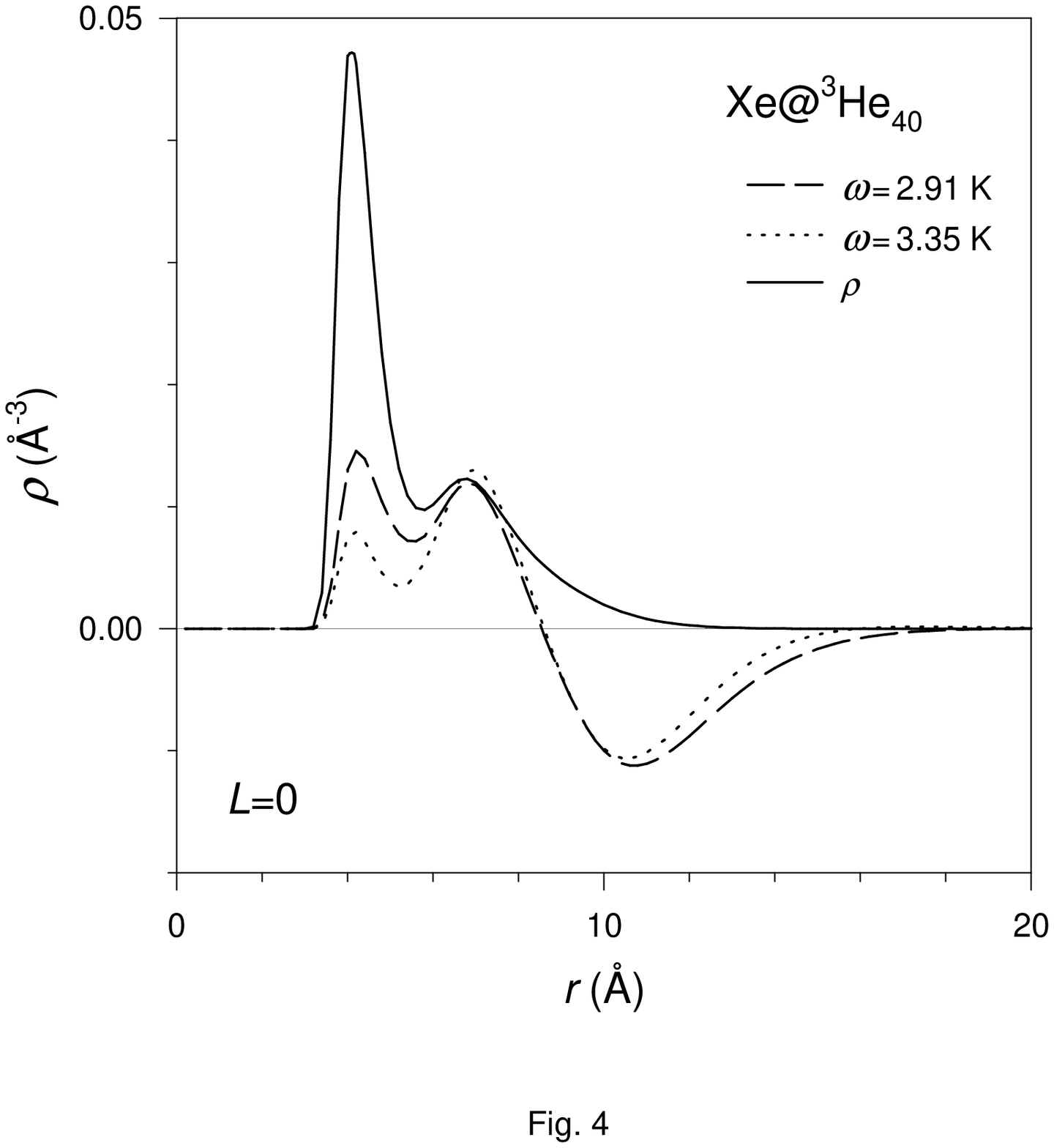}
\vspace*{13cm}
\caption{
Transition densities (arbitrary scale) corresponding to the more
intense monopole states of Xe$+^{3}$He$_{40}$.
The ground state density $\rho(r)$ is also shown.
The transition densities have been scaled to have a
common value at $r=10$~\AA.}
\label{fig3}
\end{figure}

We display in Fig. \ref{fig3} the transition densities corresponding
to the more intense
monopole peaks of Xe$+^{3}$He$_{40}$. This
figure shows the well-known fact that the monopole is a volume mode:
the induced densities have a node and penetrate inside the
drop. The bulk oscillations are connected
with the oscillations in the drop density $\rho(r)$,
also shown in Fig. \ref{fig3}, which are due to the distribution of $^3$He atoms
in solvation shells around the dopant on the one hand, and to  the
repulsive core of the effective interaction, on the other hand.

\subsection{Dipole mode}

\begin{figure}[h]
\includegraphics{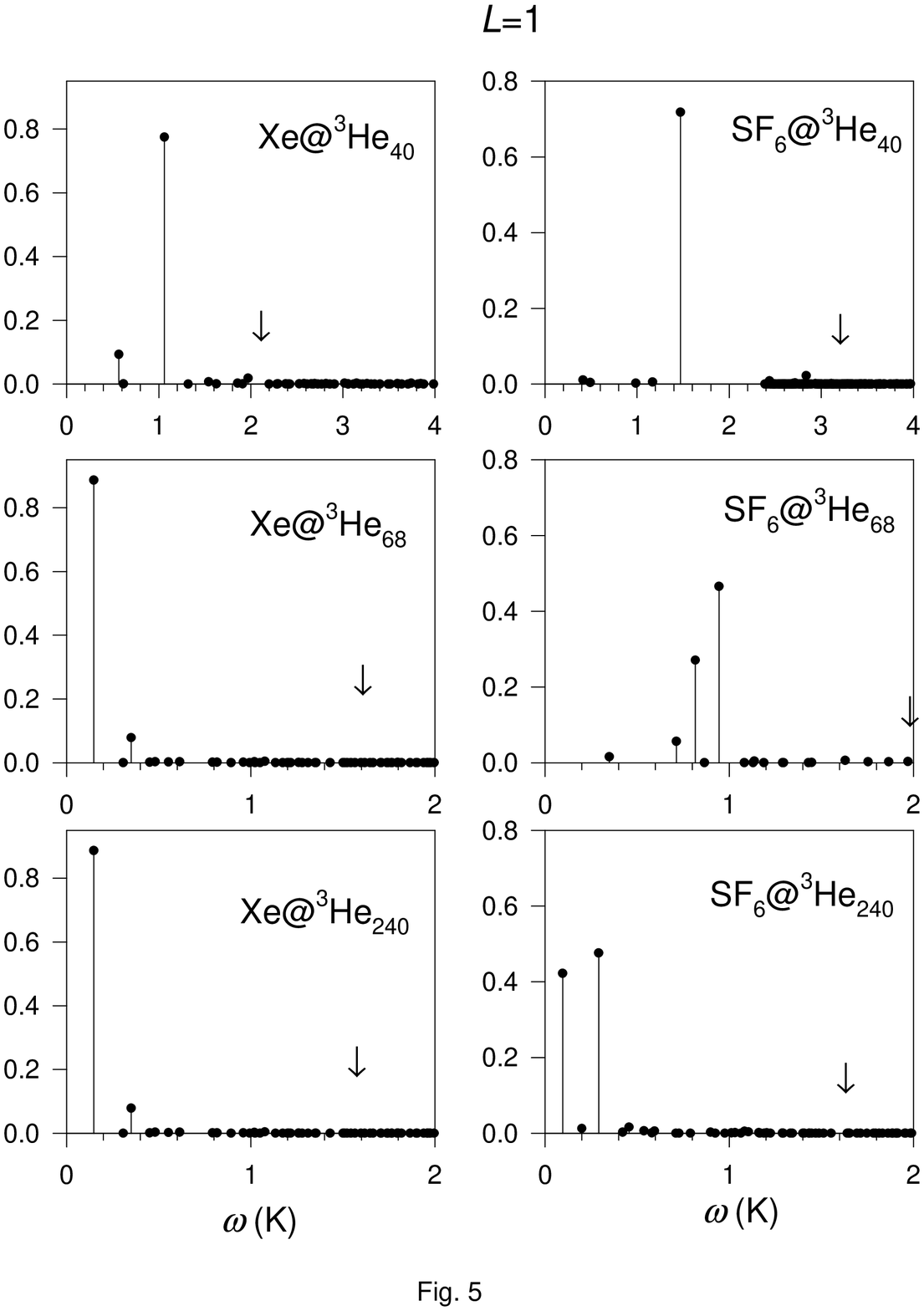}
\vspace*{18cm}
\caption{Same as Fig. \ref{fig1} for the dipole spectrum of doped drops.}
\label{fig4}
\end{figure}

The $L=1$ spectrum shows again that fragmentation increases for
the more attractive impurities (see Fig. \ref{fig4}).
In this case the mean dipole energy always lies
below the continuum threshold and decreases with the number of
$^{3}$He atoms of the drop. In spite that
small doped drops are stable to dipole fluctuations
since a large energy is needed to induce the oscillation of the
impurity against the $^3$He atoms, Fig. \ref{fig5} shows that
when the drop size increases the dipole mean energy rapidly
decreases and the mode eventually becomes unstable.
This is considered a clear signature that the impurity is
delocalized in the bulk of the drop \cite{Chi95,Ga97,Ba98}.
The dipole mode has also been found to be unstable\cite{Ga97}
for large $^4$He$_N$ drops doped with
inert atoms and SF$_6$, for which the mean dipole
energy lies in the discrete part of the spectrum.

\subsection{Quadrupole mode}

\begin{figure}[h]
\includegraphics{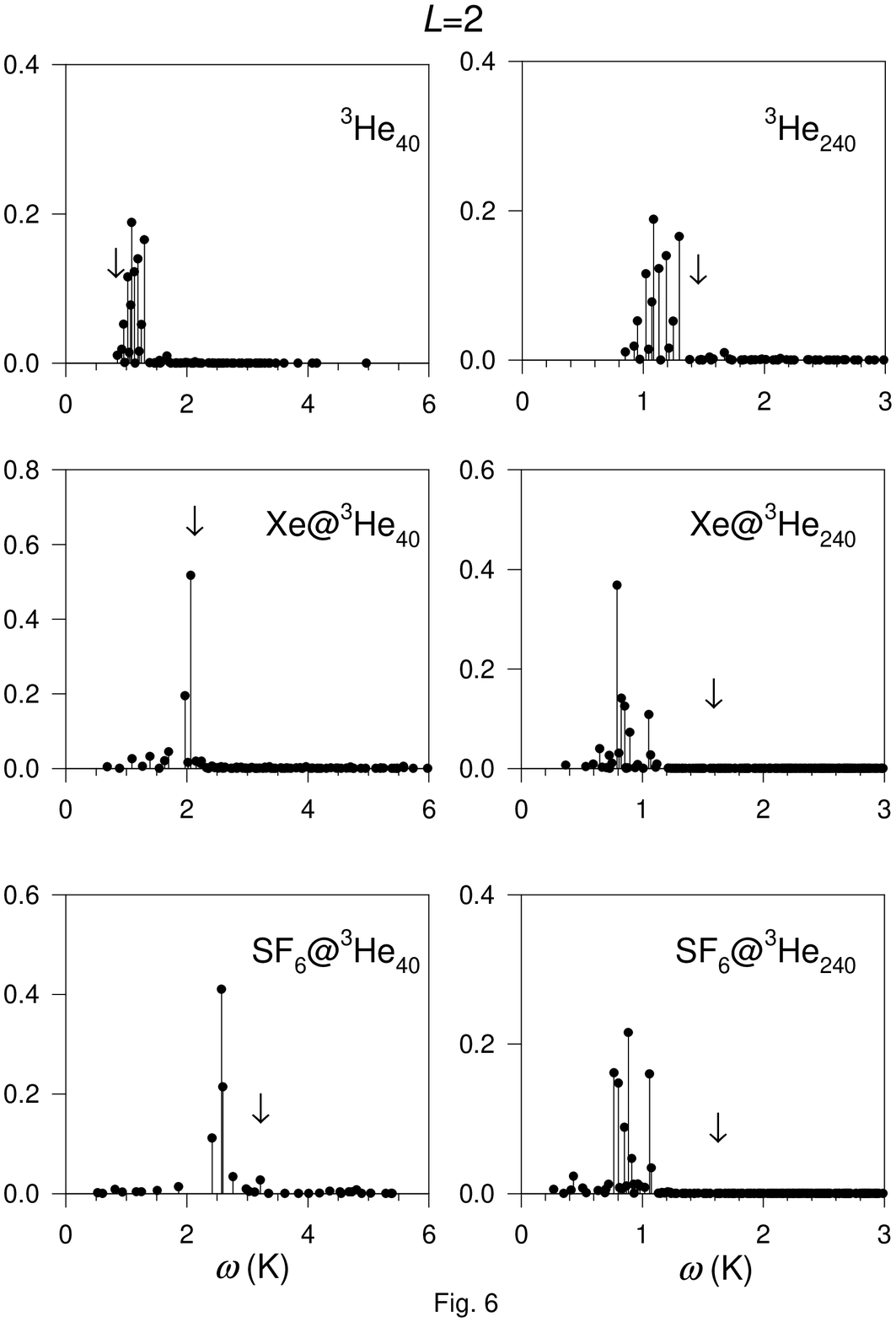}
\vspace*{18cm}
\caption{Same as Fig. \ref{fig1}. for the quadrupole spectrum.}
\label{fig6}
\end{figure}

Fig. \ref{fig6} shows that, as compared to the pure case,
the presence of a rather attractive
impurity pushes this mode downwards in energy.
When this causes the quadrupole mode to move from the continuum
to the discrete part of the energy spectrum,
the fragmentation decreases and the peak becomes more collective. This
is the case for $N = 40$, for example. For larger drops, the quadrupole
mode is below the atom emission threshold, see Fig. \ref{fig5} (this also happens
in pure drops), and the effect is not so clearly seen.

\begin{figure}[h]
\includegraphics{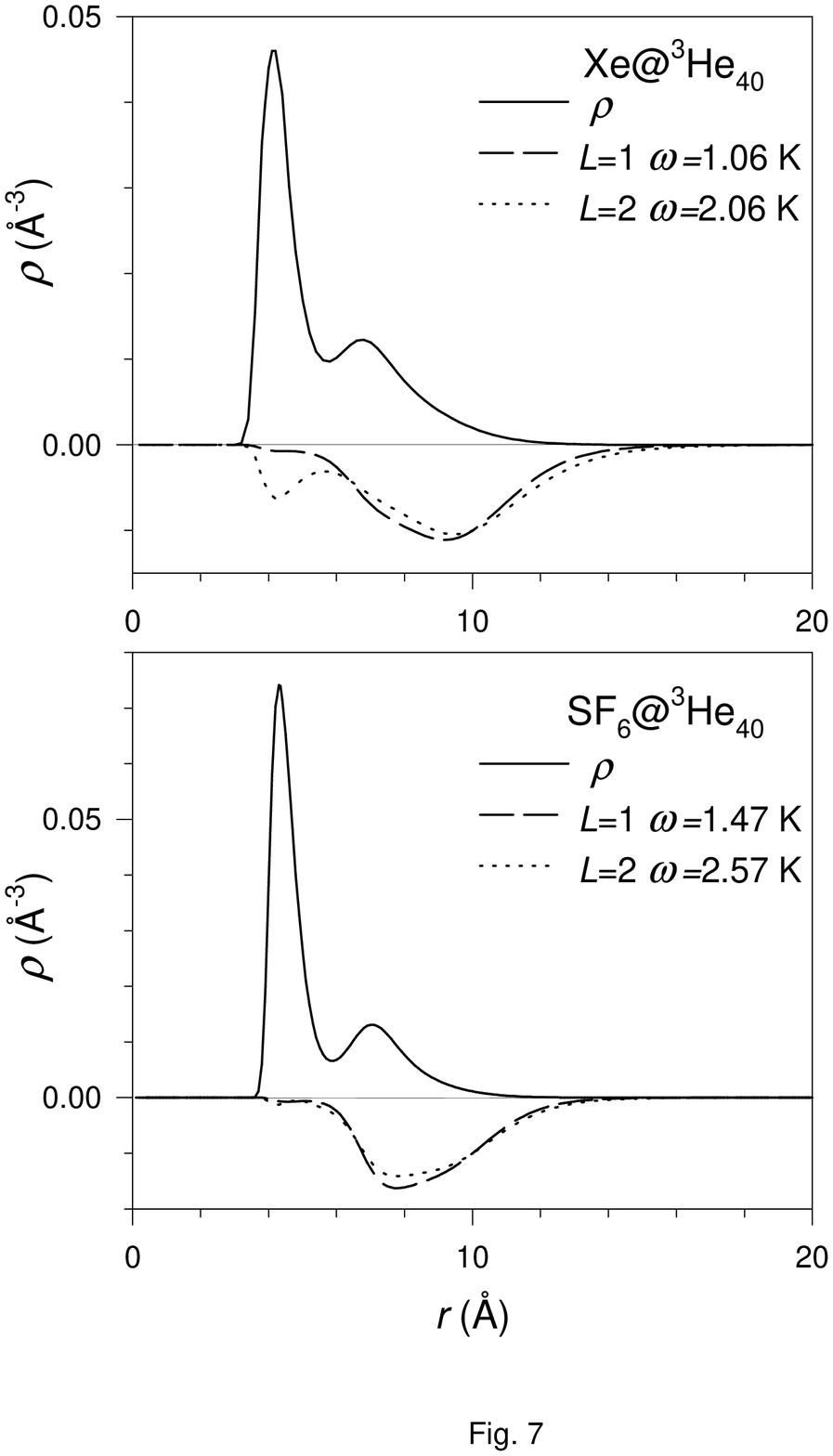}
\vspace*{18cm}
\caption{
Transition densities (arbitrary scale) corresponding to the more
intense $L=$ 1 and 2 peaks and
ground state density $\rho(r)$ of the $^{3}$He$_{40}$ drop doped
with Xe and SF$_6$.
Scaling factors as in Fig. \ref{fig3} have been used.}
\label{fig7}
\end{figure}

Examples of induced densities for dipole and quadrupole modes
are shown in Fig. \ref{fig7}.
They are  localized at the drop surface, as it corresponds to
the surface character of these modes.

We would like to close this Section indicating that in the case of
pure $^3$He droplets, a comparison with results for $L=0$ and 2
modes obtained using density functionals built using fairly
different strategies\cite{Se91,We92} yields an overall good
agreement.

\section{Conclusions}

We have investigated the multipole collective excitations of
$^{3}$He$_{N}$ drops doped with Xe atoms and SF$_6$ molecules
in the framework of the FRDF theory plus the RPA.
A  comparison with the results for pure drops shows
that the presence of these strongly attractive
impurities increases the spectrum  fragmentation.
This effect appears in volume and surface modes as well, and it
is more marked for small clusters and more attractive impurities.

The presence of an attractive impurity decreases the mean energy
of surface modes
as in the case of doped $^4$He$_N$ clusters \cite{Ga97}.
 For large clusters the mean energy of surface modes lies 
 below the atom emission threshold, whereas for the monopole 
 volume mode it is always above the threshold.

When the cluster size increases the dipole mean excitation
goes to zero, indicating that the impurity is delocalized in the
bulk of the drop for $^3$He clusters doped with Xe and SF$_6$
impurities. A similar effect was found in $^4$He clusters. From 
the experience gathered in the case of $^4$He clusters,
we may conclude that
whereas the precise value of the (rather fragmented) 
collective modes may be sensitive
to the arbitrariness introduced in the choice of some of the FRDF 
ingredients, as for example the core of the
screened Lennard-Jones potential, we consider robust 
the prediction of the impurity delocalization, as well as the
evolution of the mean mode energies with the number of atoms.

\acknowledgments

This work has been supported in part by DGESIC (Spain), grants
PB98-0124 and PB98-1247, and by the Generalitat de Catalunya
Program 2000SGR-00024.
\appendix
\section{}

For a spherically symmetric system the dimension of the ph space
can be drastically reduced by analytically summing over
the degenerate third components of the angular momentum.
We illustrate this point taking as an example
the Lennard-Jones contribution
$V_{LJ}(\vec r\,)$ to the ph interaction assuming that the
ph states are coupled to yield an orbital angular momentum
($L,M_L$), and a spin ($S,M_S$).
Using the appropriate sp quantum numbers we represent
the orbital $\phi_a$ in  coordinate and spin spaces as

\begin{equation}
\phi_a\equiv\phi_{a\ell_a\mu_a}(\vec r\,)\chi_{{1\over2}\sigma_a} \ .
\end{equation}
The matrix elements of the residual interaction (\ref{vpht}) between the
ph states can then be obtained as

\begin{eqnarray} \label{A1}
\langle mj \vert V_{ph} \vert ni \rangle &=& \sum_{\sigma{\rm 's}}
(-1)^{\frac{1}{2}-\sigma_i+\frac{1}{2}-\sigma_j}\,
(\frac{1}{2}\frac{1}{2}\sigma_m-\sigma_i|S M_S)\,
(\frac{1}{2}\frac{1}{2}\sigma_n-\sigma_j|S M_S)\,
\langle\chi_{\sigma_m}| \chi_{\sigma_i} \rangle\,
\langle\chi_{\sigma_j}| \chi_{\sigma_n} \rangle  \times
\nonumber\\
&&\sum_{\mu{\rm 's}} (-1)^{\ell_i-\mu_i+\ell_j-\mu_j}\,
\left(\ell_m\ell_i\mu_m-\mu_i|L M_L \right)\,
\left(\ell_n\ell_j\mu_n-\mu_j|L M_L \right)  \times \nonumber \\
&&\langle \phi_{m\ell_m\mu_m}\phi_{j\ell_j\mu_j}| V_{ph}(\vec r_{12}) |
\phi_{n\ell_n\mu_n}\phi_{i\ell_i\mu_i}\rangle \ .
\end{eqnarray}
Due to the spherical symmetry of the mean field, the sp wave functions
separate in radial and angular components:
\begin{equation}\label{A2}
\phi_{a\ell\mu}(\vec r\,) =
\frac{u_{a\ell}(r)}{r}\, Y_{\ell\mu}(\hat{r})\ ,
\end{equation}
and one can perform a multipole expansion of the ph interaction
\begin{equation}
\label{A3}
V_{ph}(r_{12}) = \sum_{LM}{
V_L(r_1,r_2)\, Y^*_{LM}(\hat{r}_1)\, Y_{LM} (\hat{r}_2)
}\ .
\end{equation}

The expression for the Lennard-Jones term of Eq. (\ref{vpht})
can be finally written in terms
of the reduced matrix elements of the spherical harmonics\cite{Con80}
as

\begin{equation}\label{A4}
\langle mj|V_{LJ}|ni\rangle = 2\delta_{S0} \frac{1}{\sqrt{2L+1}}\,
\langle\ell_m\vert\vert Y_L \vert\vert \ell_i\rangle\,
\langle\ell_n\vert\vert Y_L \vert\vert \ell_j\rangle\,
I^0_{mjni}  \ ,
\end{equation}
where $I^0_{mjni}$ is the radial integral

\begin{equation}\label{A5}
I^0_{mjni} =\int_0^{\infty} dr_1 \int_0^{\infty} dr_2\,
u_{m\ell_m}(r_1)\, u_{j\ell_j}(r_2)\,
V_L(r_1, r_2)\,
u_{n\ell_n}(r_2)\, u_{i\ell_i}(r_1)\ .
\end{equation}
These integrals are obtained numerically.
To describe the states above the continuum
threshold we have followed the usual prescription of enclosing
the system in a sphere of large radius
and require that the radial wave functions vanish at
this distance. In this way we obtain a discrete spectrum of states
that replaces  the continuum.
We have checked that the results are stable against reasonable
changes of the radius of the sphere, which we have taken to be about
3 times larger than the mean square radius of the cluster.

\pagebreak

\end{document}